\documentclass[prl,twocolumn, superscriptaddress, preprintnumbers, showpacs, nofootinbib]{revtex4}

\usepackage{amsfonts,amssymb,amsmath}
\usepackage{bm,bbm}
\usepackage{epsfig}

\newcommand{\beq}{\begin{eqnarray}}
\newcommand{\eeq}{\end{eqnarray}}
\newcommand{\nn}{\nonumber}
\newcommand{\eql}[1]{\label{eq:#1}}
\newcommand{\eq}[1]{(\ref{eq:#1})}
\newcommand{\fig}[1]{Fig.~\ref{fig:#1}}

\newcommand{\gsim}{\mathrel{\lower.8ex\vbox{\lineskip=.1ex\baselineskip=0ex
                   \hbox{$>$}\hbox{$\sim$}}}}
\newcommand{\lsim}{\mathrel{\lower.8ex\vbox{\lineskip=.1ex\baselineskip=0ex
                   \hbox{$<$}\hbox{$\sim$}}}}

\newcommand{\too}{\longrightarrow}

\newcommand{\avg}[1]{\left\langle #1 \right\rangle}

\newcommand{\lt}{\left}
\newcommand{\rt}{\right}

\newcommand{\fr}[2]{\frac{#1}{#2}}

\newcommand{\del}{\partial}
\newcommand{\Cases}[1]{\left\{\begin{array}{l}#1\end{array}\right.}

\newcommand{\td}[1]{\tilde{#1}}
\newcommand{\wtd}[1]{\widetilde{#1}}
\newcommand{\wba}[1]{\overline{#1}}
\newcommand{\de}{\delta}
\newcommand{\De}{\Delta}
\newcommand{\La}{\Lambda}
\newcommand{\tht}{\theta}

\newcommand{\cO}{\mathcal{O}}

\newcommand{\U}{\mathop{\text{U}}}

\begin{document}

%%%%%%%%%%%%%%%%
%%%%%%%%%%%%%%%%

\title{A Common Origin for Neutrino Anarchy and Charged Hierarchies}

\author{Kaustubh Agashe}
\affiliation{Center for Fundamental Physics, University of Maryland, College Park, MD 20742, USA}

\author{Takemichi Okui}
\affiliation{Center for Fundamental Physics, University of Maryland, College Park, MD 20742, USA}
\affiliation{Department of Physics and Astronomy, Johns Hopkins University, Baltimore, MD 21218, USA}

\author{Raman Sundrum}
\affiliation{Department of Physics and Astronomy, Johns Hopkins University, Baltimore, MD 21218, USA}
\affiliation{Center for Fundamental Physics, University of Maryland, College Park, MD 20742, USA}

\preprint{UMD-PP-08-019}

\begin{abstract}

The generation of exponential flavor hierarchies from extra-dimensional
wavefunction overlaps is re-examined. We find, surprisingly, that coexistence
of anarchic fermion mass matrices with such hierarchies is intrinsic and
natural to this setting. The salient features of charged fermion and neutrino
masses and mixings can thereby be captured within a single
framework. Both Dirac and Majorana neutrinos can be realized. The
minimal phenomenological consequences are discussed, including the need
for a fundamental scale far above the weak scale to adequately suppress
flavor-changing neutral currents. Two broad scenarios for stabilizing
this electroweak hierarchy are studied, warped compactification and
supersymmetry. In warped compactifications and ``Flavorful Supersymmetry,'' where non-trivial flavor structure
appears in the new TeV physics, Dirac neutrinos are strongly favored
over Majorana by lepton flavor violation tests. We argue that this is
part of a more general result for flavor-sensitive TeV-scale physics.
Our scenario strongly suggests that the supersymmetric flavor problem is
not solved locally in the extra dimension, but rather at or below the
compactification scale. In the supersymmetric Dirac case, we discuss how
the appearance of light right-handed sneutrinos considerably alters the
physics of dark matter. 

\end{abstract}


\maketitle

%%%%%%%%%%%%%%%%
%%%%%%%%%%%%%%%%

The quark and charged lepton masses and the CKM matrix exhibit an
intriguing and hierarchical structure that strongly motivates the search
for an underlying mechanism. Remarkably, {\it exponential} hierarchies
in masses and mixing angles can naturally arise when Standard Model (SM)
fields propagate in extra spatial dimensions \cite{ArkaniHamed:1999dc,
Gherghetta:2000qt, Kaplan:2001ga} (see \cite{Donoghue:1997rn} and \cite{Grossman:1999ra} for related ideas).  
Most simply, consider a field $\phi$
in a flat extra dimension ($0 \leq y \leq a$) with 5D mass parameter
$M$. The observed particles correspond to 4D massless (or ``zero-'') modes, 
$\phi=\phi^{(0)}(y) \, e^{ip\cdot x}$, with the 4D momentum satisfying $p^2=0$.  $\phi^{(0)}$ then satisfies
\beq
  (\Box_{\rm 5D} + M^2) \phi^{(0)}  e^{ip\cdot x} 
  =  e^{ip\cdot x} (-\del_y^2 + M^2) \phi^{(0)}
  = 0  \,,
\eql{Klein-Gordon}
\eeq
subject to boundary effects at $y = 0, a$ \cite{Gherghetta:2000qt,
Georgi:2000wb}.  Such effects generically cause one of the two
exponential solutions to dominate:
\beq
  \phi^{(0)}(y) \sim e^{-My} \quad \text{or} \quad e^{+M(y-a)}  \,.            
\eql{exponentials}
\eeq
These translate to exponential hierarchies among the 4D effective Yukawa
couplings via wavefunction overlaps in matching to the 5D theory.

Naively, this also suggests hierarchical masses and mixings among
neutrinos. Of course, recent data, with $O(1)$ mixing angles and
ratios of mass-splittings, is in conflict with this expectation.  
The coexistence of hierarchical and non-hierarchical mass matrices for 
closely related particle species is quite non-trivial to explain theoretically.
Extra structures and symmetries have been invoked to accommodate 
this surprising feature \cite{Mohapatra:2006gs}.

However, in this paper, we will show that closer inspection of the
relevant wavefunction overlaps reveals a more interesting range of
possibilities which can elegantly capture the apparent ``switching''
behavior in going from hierarchical charged fermions to
non-hierarchical but light neutrinos. Such mechanisms exist for both
Dirac and Majorana neutrinos, although they take quite different forms.
Our unified treatment of charged fermions and neutrinos has striking
implications for well-known approaches to the hierarchy
problem such as supersymmetry, warped extra dimensions and strong
dynamics.

\begin{figure}
\epsfig{file=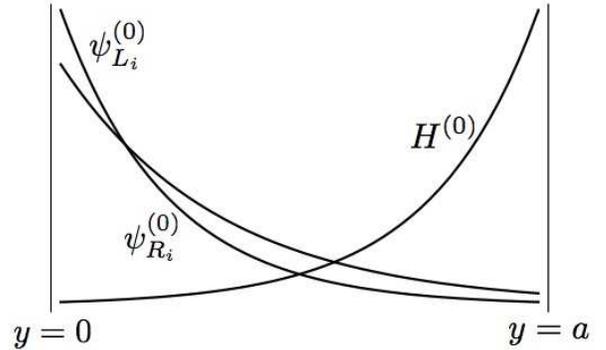, width=0.9\linewidth}
\caption{The case where both fermion chiralities lean away 
from $H$.}
\label{fig:profiles}
\end{figure}

Let us see how flavor hierarchies emerge from the exponentials
\eq{exponentials}. We assume no large hierarchies among fundamental 5D
parameters, and will try to generate large hierarchies in the 4D
effective theory purely via exponentiation $\sim e^{-Ma}$.  We first
study Dirac neutrinos so that all fermion masses can be treated
uniformly. 5D boundary and bulk Yukawa couplings induce effective 4D
Yukawa matrices,
\beq
  Y_{{\rm 4D},ij} 
  = \int_0^a \!\! dy\, Y_{{\rm 5D},ij} (y) \, 
    H^{(0)}(y) \psi_{L_i}^{(0)*} (y) \psi_{R_j}^{(0)}(y)  \,,
\eql{integral}
\eeq
where $Y_{{\rm 5D},ij}(y) \equiv Y_{0,ij} \de(y) + Y_{{\rm bulk},ij} +
Y_{a,ij}\de(y-a)$. Each of the zero-modes $H^{(0)}$, $\psi_{L_i}^{(0)}$,
$\psi_{R_j}^{(0)}$ satisfies \eq{exponentials} (up to non-exponential
normalization factors), where $M$ and the sign of exponent can depend on
both SM representation ($q_L$, $u_R$, $d_R$, $\ell_L$, $e_R$, $\nu_R$,
$H$) and generation $i=1,2,3$.

Now, note that the integral \eq{integral} is generically exponentially
dominated at $y \sim 0$ or $\sim a$. For example, consider a case where
both $\psi_{L_i}^{(0)}$ and $\psi_{R_j}^{(0)}$ lean away from $H^{(0)}$
as in \fig{profiles}, and imagine an assortment of $M_{L_i}$ and
$M_{R_j}$.  There are two cases, the integral
\eq{integral} being dominated at $y \sim 0$ or $\sim a$, depending on
whether $M_{L_i}+M_{R_j} > M_H$ or $< M_H$, respectively:
\beq
 && Y_{{\rm 4D},ij}
    \sim \int_0^a \! dy \, 
         Y_{{\rm 5D},ij} (y) \, e^{-(M_{L_i}+M_{R_j})y + M_H (y-a)}  \nn\\
 && \hspace{-1ex}
    {}^{(M_{L_i}+M_{R_j} > M_H)}
    \hbox{\Huge $\swarrow$} \qquad \hbox{\Huge $\searrow$}
    {}^{(M_{L_i}+M_{R_j} < M_H)}  \eql{bifurcation}\\
 && \;\sim \wtd{Y}_{0,ij} \, e^{-M_H a}  
    \qquad\> \hbox{\LARGE $\ll$} \quad 
    \sim \wtd{Y}_{a,ij} \, e^{-(M_{L_i}+M_{R_j}) a} \,,\nn
\eeq
where $\wtd{Y}_{0,ij}$ ($\wtd{Y}_{a,ij}$) is an $O(1)$ linear
combination of $Y_{{\rm bulk},ij}$ and $Y_{0,ij}$ ($Y_{a,ij}$). The
strong inequality in the last line of \eq{bifurcation} follows simply from
the condition on the exponents in the $y \sim a$ case, $M_{L_i}+M_{R_j} < M_H$.  The
$y\sim a$ case is the classic model of charged fermion mass matrices,
yielding exponential hierarchies in masses and mixings
\cite{Gherghetta:2000qt, Kaplan:2001ga}. Note that this would be the
only case if $H$ were boundary-localized ($M_H \to \infty$) as is often
assumed in extra-dimensional flavor models. But if that were true, it
would strongly suggest that $\nu$ masses and mixings should exhibit
hierarchies comparable to charged fermions in stark contrast to data,
unless there is some extra rationale for degeneracies among the
$M_{\ell_{Li}}$ and the $M_{\nu_{Rj}}$.

On the other hand, for a generic bulk $H$ ($M_H < \infty$), we can
elegantly accommodate $\nu$ data while simultaneously capturing charged
fermion hierarchies.%
\footnote{Even if $H$ is exactly boundary-localized, at the quantum level, 
loops containing fermion pairs with Higgs quantum numbers reproduce the 
effects of a bulk Higgs.}
For sufficiently large $M_{\nu_{Rj}}$ such that
$M_{\ell_{Li}} + M_{\nu_{Rj}} > M_H$, the $\nu$'s switch to the $y\sim 0$
case by \eq{bifurcation}, which naturally has no large {\it flavor-dependent} hierarchies. Note that this does not affect our discussion above regarding
charged fermion hierarchies generated at $y \sim a$.
Moreover, the last line of \eq{bifurcation} means that the $\nu$'s are 
exponentially lighter than the charged fermions.  
The structures and relations between the $\nu$ and charged mass matrices
are quite robust because they
derive from the branching in \eq{bifurcation}, based on simple inequalities
among the $M$'s.

Returning to \eq{integral}, we could also consider $\psi_{L,R}^{(0)}$
leaning {\it toward} $H^{(0)}$ in contrast to \fig{profiles}.  Indeed,
an $O(1)$ top Yukawa coupling does not match either exponentially
suppressed case in \eq{bifurcation}, but robustly follows once
$t_{L,R}^{(0)}$ both lean towards $H^{(0)}$. Then different leanings for
$\psi_{L,R}^{(0)}$ contribute to a smaller bottom quark mass and mixing
angles. In this way the simple extra-dimensional framework captures the
presence and absence of hierarchies across the range of flavor physics.
 
The case of Majorana neutrinos works differently. In this case, the
smallness of $m_\nu$ comes just from its non-renormalizable origins,
$\ell_L \ell_LHH/\La$.  As for the non-hierarchical
nature of neutrinos, it is a generic consequence precisely when all
$\ell_{L_i}^{(0)}$ lean toward $H^{(0)}$, regardless of the precise
$M_{\ell_{Li}}$:
\beq
  m_{\nu,ij} 
  \sim  O(1)_{ij} \, \fr{v^2}{\La} \,,
\eql{Majorana}
\eeq
where $v$ is the weak scale.
The hierarchical
structure among charged leptons can be generated if $e_{R_i}^{(0)}$ lean
away from $H^{(0)}$. 

The minimal experimental implications reduce in the $\nu$ sector to
those of the ``neutrino mass anarchy'' scenario \cite{Hall:1999sn},
namely, $\tht_{13}$ should be close to the current upper bound $\sim 0.2$, 
and CP-violating phase(s) should be $O(1)$.  More
speculatively, at the other end of the spectrum, a 4th SM generation is
a natural possibility in the Dirac $\nu$ case, and in order for it to be
heavy, their $\psi_{L,R}^{(0)}$ must be leaning toward $H^{(0)}$, just
like the top.  Thus, we expect large mixing with the 3rd generation,
which would dominate the phenomenology.

We must also consider the impact of Kaluza-Klein (KK) excitations. While
4D gauge fields couple flavor-blindly by 4D gauge invariance, gauge KK
modes are sensitive to the flavor-dependent profiles of the fermions. 
Exchanging them will generate flavor-violating 4-fermion operators with 
strength $\sim g_{\rm SM}^2 /M_{\rm KK}^2$ with $M_{\rm KK} \sim a^{-1}$. 
To avoid excessive flavor-changing
neutral currents (FCNCs) from such interactions we need $a^{-1} \gsim
1000$ TeV \cite{Delgado:1999sv}. Therefore, we should address the
hierarchy problem between the electroweak scale and at least this high
scale.  We will consider two solutions, warping the
above extra dimension \cite{Csaki:2004ay} and supersymmetry (SUSY)
\cite{Martin:1997ns}.

The simplest 5D warped spacetime is given by:
\beq
  ds^2 = e^{-2ky} dx_{\rm 4D}^2 - dy^2  \qquad (0 \leq y \leq a)  \,.
\eeq
The curvature scale $k$ is comparable to the typical 5D mass scale, such
as $M$ and $a^{-1}$, and is taken to be very high.  An exponentially
small $v \sim k e^{-ka}$ emerges naturally for $a \gsim
k^{-1}$, provided that $H^{(0)}$ is sufficiently localized near $y=a$
\cite{Randall:1999ee}.  The zero-mode profiles continue to be
exponentials in $y$, but with the curvature-modified exponents
\cite{Gherghetta:2000qt}:
\begin{align}
  e^{M_H (y-a)} 
     &\too e^{\bigl( k+\sqrt{4k^2 + M_H^2} \bigr) (y-a)}
            \text{~~for $H^{(0)}$,}  \nn\\
  \Cases{e^{-My} \cr e^{M(y-a)}}
     &\too \Cases{e^{(k/2-M) y}  \cr e^{(k/2+M) (y-a)}}
            \text{for each $\psi^{(0)}$.}
\eql{warped-exponents}
\end{align}
The flavor mechanisms described earlier in flat spacetime transfer to
the warped setting with these modifications. The curvature-modified
stability bound, $M_H^2 \geq -4k^2$ \cite{Breitenlohner:1982bm}, has the
following important consequence for our Dirac $\nu$ scenario. Plugging
\eq{warped-exponents} into \eq{bifurcation} we see that the light
non-hierarchical fermions have Yukawa couplings $\lsim e^{-ka}$, the
size of the warped hierarchy itself:
\beq
  Y_{\text{non-hierarchical}} \lsim \fr{v}{k}  \,.
\eql{maxY}
\eeq
It is intriguing that at the bound, the observed neutrino data with
$m_{\nu} \sim 0.1$ eV then imply a rough hierarchy from $k \sim 10^{15}$
GeV down to the weak scale.

FCNCs mediated by KK excitations are more subtle due to warping
\cite{Gherghetta:2000qt}. This is because even when all 5D fundamental
scales are very large, low-lying KK masses are exponentially light,
$M_{\rm KK} \sim k e^{-ka} \sim$ TeV.  However, these excitations have
extra-dimensional profiles concentrated near $y=a$.  Therefore, if the
fermions are configured as in \fig{profiles}, the wavefunction overlaps
suppress the flavor-dependent couplings of these KK
excitations to SM fermions, similarly to the 4D Yukawa couplings.  
Such factors can sufficiently suppress hadronic flavor violation
for $M_{\rm KK} \sim$ several TeV \cite{Huber:2003tu}.

Let us now turn to leptons, where $\nu$ data force us to consider FCNCs
among charged leptons such as $\mu \to e\gamma$ and $\mu \to e$
conversions in nuclei. Our scenario for Dirac neutrinos is quite
analogous to the quark sector. FCNCs are suppressed by wavefunction
overlaps dominated at $y \sim a$, comparable to {\it charged} lepton
Yukawa couplings.  Again, $M_{\rm KK} \sim$ several TeV is sufficient to
satisfy the current bounds \cite{Agashe:2006iy}. Because of the
switching behavior \eq{bifurcation}, this physics is divorced from the
generation of large $\nu$ mixing angles at $y\sim 0$.

Contrast this to the case where $H$ is exactly
boundary-localized ($M_H \to \infty$), so that there is no switching
behavior. As in flat spacetime, this case would then lose our
explanation for non-hierarchical $\nu$ structure. Furthermore, the $\nu$
mass matrix is then necessarily generated at $y=a$, which exacerbates
charged lepton FCNCs due to the large $\nu$ mixings \cite{Perez:2008ee}. 

Our scenario for Majorana $\nu$'s is completely excluded when warping
solves the hierarchy problem. Recall that in this scenario, $\ell_{L_i}^{(0)}$
lean in the same direction as $H^{(0)}$ and therefore the same direction
as the KK modes. Consequently, KK-mediated FCNC operators such as
$\bar{\mu} e \bar{e} e / M_{\rm KK}^2$ are not suppressed by small
wavefunction overlaps. Bounds on $\mu \to 3 e$ would then require KK
modes heavier than $\sim O(100)$ TeV, grossly at odds with their solving
the hierarchy problem.

The successful implementation of our flavor mechanism in warped
compactifications extrapolates to a larger class of purely 4D but
strongly coupled theories via the AdS/CFT correspondence
\cite{Aharony:1999ti}.  The large number of states implied by the extra
dimension are reproduced by the many composites of a strong sector which
is conformal over a large hierarchy. The 4D theory also contains
elementary fermions with SM quantum numbers, $\psi_{L,R}$, outside of
the strong sector. The mechanism of partial compositeness (PC)
\cite{Kaplan:1991dc} introduces couplings, $\psi_{L,R} \, {\cal
O}_{R,L}$, to fermionic operators of the strong sector. Electroweak
symmetry breaking (EWSB), assumed to occur in the strong sector at scale
$v$, is communicated to the $\psi_{L,R}$ via these couplings. This
naturally generates hierarchical Dirac fermion mass matrices via running
down from a high fundamental scale $\La$:
\beq
  m_{ij} \sim O(1)_{ij}\, v \lt( \fr{v}{\La} \rt)^{\De_{L_i} + \De_{R_j} - 5},
\eql{PC}
\eeq
where $\De_{L_i, R_j}$ are scaling dimensions of ${\cal O}_{L_i,R_j}$. 
This is the dual of the right-hand branch of \eq{bifurcation} after the
curvature modification of \eq{warped-exponents}, where $\De - 2
\stackrel{{\rm AdS/CFT}}{ \longleftrightarrow } M$.

Let us turn to the dual of the left-hand branch of \eq{bifurcation}.
There must also be composite operators with Higgs quantum numbers (for
example, ${\cal O}_L {\cal O}_R$). Let ${\cal O}_H$ be the operator of
this type with the lowest scaling dimension, $\De_H$. Then elementary
fermions can couple to it bilinearly, $\wba{\psi}_{L_i} \psi_{R_j} \cO_H
/ \La^{\De_H - 1}$, leading after EWSB to
\beq
  m_{ij} 
  \sim O(1)_{ij} \fr{\avg{\cO_H}}{\La^{\De_H - 1}}  
  \sim O(1)_{ij}\, v \lt( \fr{v}{\La} \rt)^{\De_H - 1}  \,.
\eql{ETC}
\eeq
In this way, such {\it anarchic} contributions to fermion masses follow
inevitably from the fermion hierarchies generated by PC. 
In addition, in strong sectors of large-$N$ type, the gauge singlet operator 
${\cal O}_H^{ \dagger } {\cal O}_H$ has dimension $\sim 2 \De_H$. The
IR-stability of the strong conformal dynamics against perturbations of
this type then requires $2 \De_H \gsim 4$. This implies that fermions
masses from \eq{ETC} are very small, $m_{ij} \lsim v \lt( v / \La \rt)$,
which is the dual of \eq{maxY}.  Clearly, for $\De_{L_i} + \De_{R_j} <
\De_H + 4$, PC \eq{PC} dominates, whereas for $\De_{L_i} + \De_{R_j} >
\De_H + 4$, the $\avg{\cO_H}$-induced fermion masses \eq{ETC} dominate.
This bifurcation is the dual of \eq{bifurcation} (with the modification
\eq{warped-exponents}).

Effects like \eq{ETC} appear in walking technicolor theories
from integrating out extended technicolor (ETC) physics at $\La$, where 
Yukawa hierarchies follow from a hierarchy of different $\La$'s
\cite{Hill:2002ap}.  While the PC and ETC mechanisms have been
separately discussed in the literature, the novelty of
our framework is that one implies the other, the switching behavior
arising out of their competition.

We now turn to the supersymmetric solution for stabilizing the large
electroweak hierarchy $a^{-1} \gg v$, but in {\it flat} spacetime again.
In 5D supersymmetry it is difficult to have non-zero $Y_{\rm bulk}$, but
boundary-localized superpotentials are straightforward. Our
flavor mechanisms proceed as before but with $\wtd{Y}_{0,a}$ in
\eq{bifurcation} now given by just the boundary couplings $Y_{0,a}$.

The nature of the SUSY flavor problem is somewhat transformed by the
extra-dimensional flavor structure. The simplest ansatz for weak-scale
soft masses that satisfies FCNC constraints is squark and slepton
universality. In purely 4D, there are two options: either (A) universal
soft masses emerge from some special mechanism of Planck scale physics,
unspoiled by the modest flavor violation of MSSM loop effects in the IR,
or (B) universality of soft masses is due to an effective field theory
(EFT) mechanism in the IR. But in our 5D picture,
option (A) is implausible. While there is a 4D MSSM regime below the
compactification scale $a^{-1}$, with flavor violation suppressed by
wavefunction overlaps, the 5D regime between the 5D Planck scale and
$a^{-1}$ has maximal $O(1)$ flavor-violating interactions and hence
loops. Therefore, we are left with option (B).

4D gauge and gravity effects, that are non-local from the 5D
perspective, can naturally provide such IR-dominated mechanisms. Gauge
mediation (GM) \cite{Giudice:1998bp}, gaugino mediation ($\td{g}$M) 
\cite{Kaplan:1999ac}, anomaly mediation (AM) \cite{Randall:1998uk}, and 
(D-term dominated) gravity-loop mediation \cite{Gregoire:2005jr} are 
prime examples. These mechanisms certainly do solve the SUSY flavor problem, 
but unfortunately then the physics of the superpartners will not contain 
any experimental traces of the extra-dimensional origins of flavor.

But flavor universality of soft terms need not be exact. One can also
have contributions to soft terms which are sensitive to 5D flavor
structure, as long as they are subdominant enough to satisfy low-energy
flavor tests \cite{Kaplan:2000av} \cite{Kaplan:2001ga}.  
An interesting example is the
scenario of ``Flavorful Supersymmetry'' \cite{Nomura:2007ap}.  This can
arise in $\td{g}$M and AM scenarios in which
flavor-blindness depends on {\it sequestering,} the separation in extra
dimensions of SM matter from the SUSY-breaking sector.  Approximate
sequestering occurs in the setup of \fig{profiles} if SUSY-breaking is
localized at $y=a$, where it has small overlap with the zero-modes of
the matter superfields.  Then, the exponential suppression factors
appearing in the non-universal corrections to soft terms are the same as
those in Yukawa couplings.

An important observation is that our Majorana $\nu$ scenario is
incompatible with this picture of Flavorful SUSY, because the
$\ell_{L_i}$ then lean towards $y=a$.  This implies a maximal breaking
of sequestering and $O(1)$ lepton-flavor violation in soft terms, which
is disfavored at the TeV scale. As in the warped case above, only our
Dirac $\nu$ scenario survives. In contrast, mediation mechanisms which
do not require sequestering (e.g.~GM) or in which it is
achieved independently of zero-mode leanings, can be fully compatible
with our Majorana $\nu$ scenario.  But, of course, in these cases soft
terms will not provide us an imprint of flavor origins.
R-symmetric supersymmetry breaking is somewhat exceptional in that the $O(1)$
lepton flavor violation of our Majorana $\nu$ scenario can be consistent 
with flavor tests, given a modest hierarchy between the slepton and gaugino 
masses \cite{Kribs:2007ac}. 

The Dirac $\nu$ case introduces new superfields $N_i \equiv (\nu_{R_i},
\td{\nu}_{R_i})$ to weak-scale physics, whose consequences we discuss
for the different mediation mechanisms. These mechanisms all generate
soft masses based on the 4D interactions of the sparticle in question. 
The danger for the sterile $\td{\nu}_R$ is that it acquires a very small
mass $\ll m_Z$, potentially spoiling the LSP dark matter scenario. In
GM, the $\td{\nu}_R$ is indeed very light, but the very
light gravitino is already problematic for LSP dark matter.  In $\td{g}$M
(or its ``flavorful'' variant), the very light $\td{\nu}_R$ is
a new problem.  But adding $\U(1)_{B-L}$ interactions, higgsed below the
compactification scale, generates $m_{\td{\nu}_R} \sim O(m_Z)$,
making a $\td{\nu}_R$ a viable WIMP or super-WIMP \cite{Feng:2003xh}
dark matter candidate. In AM, $m_{\td{\nu}_R} \sim
O(m_Z)$ can be generated by $O(1)$ renormalizable superpotential
couplings, such as $N^3$, again leading to a $\td{\nu}_R$ WIMP. Finally,
gravity-loops at the compactification scale \cite{Gregoire:2005jr} can
combine with AM, leading to a different LSP from
$\td{\nu}_R$ that can be a viable and detectable dark matter candidate. 

We have studied the impact of our extra-dimensional flavor mechanisms on
quite different scenarios for TeV physics. In those in which the
TeV physics contains clues to the origins of flavor structure, the
Majorana $\nu$ option was disfavored by lepton-flavor violation
constraints. In fact, this is more general. The flavor-violating
exponential suppression factors, which are the building blocks of
Yukawa hierarchies in our 5D picture, can reappear in TeV physics
whenever this new physics is localized near $H^{(0)}$. While TeV physics with
maximal flavor violation would typically induce too large FCNCs at low
energies, these exponential overlaps can suppress them to acceptable
levels.  However, for our Majorana $\nu$ scenario, these suppression
factors are absent since the $\ell_{L_i}$ and the new physics (and the
Higgs) are localized in the same region of the extra dimension.
Therefore, in this type of flavor-sensitive TeV scenario, Dirac neutrinos 
are preferred.

We thank Neal Weiner for useful comments. 
We also thank one of our referees for enlightening comments on 4D models
with ``switching'' behavior.
This work was supported in part by NSF grants PHY-0401513, PHY-0652363,
DOE grant DE-FG02-03ER4127, the Alfred P.~Sloan Foundation, 
the Johns Hopkins Theoretical Interdisciplinary Physics and Astrophysics Center,
and the Aspen Center for Physics.

%%%%%%%%%%%%%%%%
%%%%%%%%%%%%%%%%

\end{document}